\documentclass[preprint2]{aastex6}

\usepackage{graphicx}
\usepackage{hyperref}
\usepackage{natbib}
\usepackage{color}
\usepackage{mathtools}
\usepackage{aas_macros}

\bibpunct{(}{)}{;}{a}{}{,}

\shorttitle{}
\shortauthors{Adhikari et al.}
\slugcomment{Accepted for publication in The Astrophysical Journal}

\begin{document}


\title{Intermediate-line Emission in AGNs: The Effect of Prescription of the Gas Density}


\author{T. P. Adhikari\altaffilmark{1}, K. Hryniewicz\altaffilmark{1}, A. R\'o\.za\'nska\altaffilmark{1},
B. Czerny\altaffilmark{2} and G. J. Ferland\altaffilmark{3,1}}
\email{tek@camk.edu.pl}
\altaffiltext{1}{Nicolaus Copernicus Astronomical Center, Polish Academy of Sciences, Bartycka 18, 00-716, Warsaw}

\altaffiltext{2}{Center for Theoretical Physics, Polish Academy of Sciences, Aleja
Lotnikow 32/46, Warsaw, Poland} 

\altaffiltext{3}{Department of Physics and Astronomy, The University of Kentucky, Lexington, KY 40506, USA} 

\begin{abstract}

The requirement of intermediate line component in the recently
observed spectra of several AGNs points to possibility of the existence 
of a physically separate region between broad line region (BLR) 
and narrow line region (NLR). In this paper we explore the emission from intermediate line 
region (ILR) by using the photoionization simulations of the gas clouds distributed radially from the 
AGN center. The gas clouds span distances typical for BLR, ILR and NLR, and the apperance of dust 
at the sublimation radius is fully taken into accout in our model. Single cloud structure
is calculated under the assumption of the constant pressure. We show that the slope of 
the power law cloud density radial profile  does not affect the existence of ILR in major types of AGN.
We found that the low ionization iron line, Fe~II, appears to be highly sensitive for the 
presence of dust and therefore becomes potential tracer of dust content in line emitting regions. 
We show that the use of  disk-like cloud density profile computed at the  upper part of the
accretion disc atmosphere reproduces the observed properties of the line emissivities. In 
particular, the distance of H${\beta}$ line inferred from our model agrees with that 
obtained from the reverberation mapping studies in Sy1 galaxy NGC~5548.
\end{abstract}



\section{Introduction}
Over the past several decades, the properties and origin of broad line 
region (BLR) as well as narrow line region (NLR) in active galactic nuclei (AGN)
are extensively discussed in the literatures 
\citep[and references therein]
{davidson1972,Krolik1981,Netzer1990,Dopita2002,baskin2014,czerny2011,Czerny2015,Czerny2017}. 
There is a general consensus that both regions are physically 
separate and spatially located at different distances from the central 
supermassive black hole (SMBH) of AGN. The above conclusion naturally came
from the lack of any significant emission from lines with full width at
half maximum (FWHM) between $\sim$ 2000 km s$^{-1}$, typical for BLR,  and 
$\sim$ 500 km s$^{-1}$, typical for NLR \citep{Boroson1992}.

Theoretically, the lack of such emission from
gas between BLR and NLR was successfully  
explained by \citet[hereafter NL93]{Netzer1993}.  The authors calculated the 
line emission from radially distributed clouds above an accretion disk, using photoionization
computations. Each cloud was a constant density slab illuminated by the same 
mean quasar continuum shape .  
The lack of line emission was successfully achieved with the introduction of 
dust. Practically it means, that 
the dust was taken into account in photoionization calculations for clouds located further 
away from SMBH at the certain radius named sublimation radius. Closer to the nucleus
the radiation field is so strong that the dust grains cannot survive. The presence  
of dust for a given gas conditions successfully suppresses line emission, and 
the gap between BLR and NLR is naturally formed. 
However, new observations with the largest instruments give us a new look at 
those objects. 

There is a growing number of AGN which exhibit the emission lines 
with intermediate FWHM $\sim$ 700--1200 km s$^{-1}$ in their spectra 
suggesting the existence of intermediate line region (ILR) in those sources. 
\citet{Brotherton1994} have defined ILR in 15 broad UV line QSO as the second component
of BLR, located at most inner part of NLR. 

\citet{Mason1996} found  evidence for an ILR with velocity
FWHM $\sim$ 1000 km s$^{-1}$ which produces a
significant amount of both permitted and
forbidden line fluxes in ultra-soft X-ray source NLSy1 RE~J1034+396. 
H$\alpha$, H$\beta$, O~[III] observed by ISIS spectrograph La Palma. 
Detailed spectral analysis of large number of 
{\it SDSS} sources have revealed the presence of intermediate component of 
line emission with velocity width in between that of broad and narrow component \citep{hu2008a,hu2008b}. {\it SDSS} sources show ILR in H$\alpha$ and H$\beta$ lines \citep{zhu2009}, mostly for NLSy1 galaxies.

For the Sy1 NGC~4151, \citet{Crenshaw2007} identified an 
intermediate line emission component with  width FWHM $=$ 1170 km s$^{-1}$, most probably
originating between the BLR and NLR. 
For Sy1 NGC~5548, ILR was found to be located at $\sim 1$ pc, 
with smaller velocity FWHM = 680~km~s$^{-1}$ \citet{Crenshaw2009}.

Moreover, the presence of ILR in 33  galaxies with 
low ionization nuclear emission-line regions (so called LINERs) was reported
by \citet{Balmaverde2016} using HST/STIS (Hubble Space Telescope/Space Telescope Imaging Spectrograph). Since typical obscuration torus is not present in LINERs, the authors suggest that the ILR takes the form of an ionized, optically thin torus. They also suggest that
this tenuous structure is present only in LINERs because of the
general paucity of gas and dust in their nuclear regions. This also
causes their low rate of accretion and low bolometric luminosity.

Recenty, \citet[hereafter AD16]{Adhikari2016} studied the ILR by 
using the photoionization simulations of an ionized gas with dust in AGN. 
Within the framework of NL93 formalism, the authors
found that in order to expect ILR emission the density of 
gas should be high, of the order of $\sim$ 10$^{11.5}$ cm $^{-3}$ at the sublimation radius, 
which means that for each cloud located at the certain distance the density in AD16 paper 
was two orders of magnitude higher than in NL93. 
For such dense matter the gas opacities always dominates over the dust opacities in a 
region of the line formation, and the dust cannot suppress the line emission as
it was present in NL93 paper. Therefore, the usual gap between BLR and NLR can not 
be created, and ILR can be present. The above result was achieved 
for three different spectral shapes  of illuminated continuum typical for Sy1.5, Sy1 and NLSy1 AGN. 
In addition, AD16 argued that the  LINERs should also exhibit the ILR emission, due to the 
low  value of luminosity and therefore the ionization parameter in those sources. 

The aim of the paper is to investigate in details the physics of ILR. We extend our previous studies (AD16) by:
(i) performing computations for constant pressure (CP) instead of constant density (CD) cloud model, (ii) searching the influence of power-law density slope on the total line emission,
(iii) including additional emission lines as  Fe~II and C~IV, (iv) considering the disk-like cloud density profile from accretion disk atmosphere, (v) using self-consistent source luminosities 
and therefore the position of the sublimation radius.

We perform the photoionization simulations by 
publicly available numerical code {\sc cloudy} version C17.00 ~\citep{Ferland2017}, 
assuming that each cloud is under constant pressure \citep{baskin2014}. 
In the first step,   we consider the model of continuous cloud distribution above an accretion disk, following approach of AD16. Nevertheless, we vary the power law index which relates to the surface cloud density distribution within the distance from SMBH. We show,  that the existence of the ILR is not sensitive to the slope of density power-law profile considered by us with the assumption of CP cloud. The same result was obtained by \citet{adhikari2017}, in case of CD clouds. 

In the second step, for each of four sources, we compute source luminosities and 
we use those values to derive the position of sublimation radius
according to formula given by \citet{Nenkova2008}. This allows us implement dust correctly in 
our photonionization calculations. 
To achieve the physically consistent density profile, we assume that clouds are created 
from upper parts of an accretion disk. By adopting black hole masses and accretion rates of four considered sources, given in the literatures, we simulate the vertical accretion disk 
structure assuming standard \citet{SS73} disk  and transfer of radiation by diffusion approximation \citep{rozanska99}. Furthermore, we employ the radial density profile obtained by solving the vertical accretion disk structure of geometrically thin disk at optical thickness $\tau$=2/3 as described in section~ \ref{sec:real_dens_profile}. This can only be done by straight comparison of cloud density with density of an upper disk atmosphere at a given radius. This idea is very close to the development done by \citet{baskin2018} who 
associates BLR directly with accretion disk atmopshere. 

For disk-like cloud density profile, we obtained the prominent 
ILR in all four sources considered. However, the sublimation radius for LINER NGC 1097 is smaller by $\sim$ 2 orders of magnitude than rest of the sources because of its low luminosity. The density drop in disk-like density profile causes mild enhancement of 
low ionization lines (LIL), while high ionization lines (HIL) are suppressed at the density drop location. This result is in agreement with two-component BLR model presented
by \citet{collin88}. The distance inferred from the time delay of H$\beta$ in NGC 5548 taken from observations agrees with the distance at 
which the H$\beta$ line peaks in our simulated line emissivity profile.

The structure of the paper is organized as follows. 
In Sec.~\ref{sec:modpar}, we describe the numerical 
model parameters taken for the photoionization computations. Whole Sec.~\ref{sec:powlad_dens} is 
devoted to the effect of slope of the power law density profile on the computed
line emission.  In addition, the comparison of CP with CD model is
explicitly shown in subsection~\ref{sec:cpcd}, while dust sensitive line Fe~II is discussed in~\ref{sec:fe2}.
Sec.~\ref{sec:real_dens_profile} contains the results we obtained for the 
disk-like density prescription we adopted. Finally the discussion 
of line emission and conclusions are presented in Sec.~\ref{sec:lines} and Sec.~\ref{sec:conc}
respectively.
	
\section{Photoionisation simulations of ILR}
\label{sec:modpar}

In order to achieve the 
established properties of ionized gas located at different distances from SMBH,
which can be responsible for the observed broad to narrow line emission, 
we consider a distribution of clouds above an accretion disk, defined at each radial distance, $r$,
by the gas density $n_{\rm H}$ at a cloud surface,  total column density $N_{\rm H}$,
and the chemical abundances. Photoionization processes are simulated with numerical code
{\sc cloudy} version C17 \citep{Ferland2017}.  

 \begin{figure}
 \hspace{-4mm}
\includegraphics[width=0.53\textwidth]{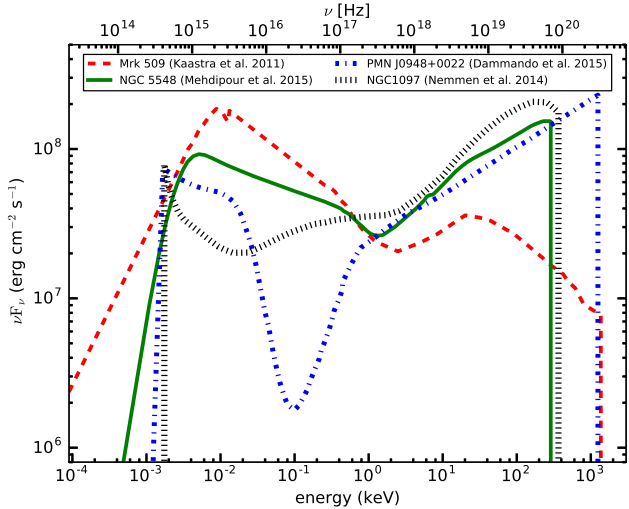}
\caption{\small Shapes of the broad band spectra used in our photoionization calculations. 
In order to see the dependence on spectral shape all SEDs are normalized to the $ L_{\rm bol} = 10\rm {^{45}~erg~s^{-1}}$. See Table~\ref{tab:param} for the exact values of luminosities.} 
\label{fig:sed_all}
\end{figure}

We used the {\sc cloudy} default solar abundances derived by \citet{Grevesse1998}, for the gas clouds being at the distance $ r \leq R_{\rm d}$, where we expect
that BLR is located. 
For the clouds at $r>R_{\rm d}$, where NLR suppose to occur, the interstellar medium 
composition (ISM) with dust grains is used \footnote{see Hazy1 {\sc cloudy} documentation for the details}. 
Depending on the radial density profile, we consider various values of sublimation radius $R_{\rm d}$, expressed in two sections below: Section ~\ref{sec:powlad_dens} and~\ref{sec:real_dens_profile}. 

In all radially distributed cloud models, we adopt the profile for the column density after NL93 and  AD16:
\begin{equation}
\label{eq:colden}
N_{\rm H}(r) = 10^{23.4} ~(r/R_{\rm d})^{-1}
\end{equation}
The normalization value of column density, $N_{\rm H} =10^{23.4}$ cm$^{-2}$, again 
is taken from NL93 and AD16 due to general agreement of the observed BLR column density.

For a given radial distance,
the cloud pressure is kept constant \citep{baskin2014}. Practically it means that the gas pressure increases with cloud optical depth as radiation pressure decreases exponentially with gas optical depth. 
This option of photoionization computations was incorporated in {\sc cloudy} as radiation pressure confinement (RPC), and used by \citet{baskin2014} for the purpose of BLR. 
The concept of RPC does not differ from the total CP models
used by \citet{rozanska2006,rozanska2008,Adhikari2015} in case of the warm absorbers in AGN. The only 
difference is in numerical treatment of the radiation pressure, which in case of second group of authors 
 is self consistently computed from the true intensity radiation field 
 \citep[see][for description of {\sc titan} code]{dumont2000}.
Nevertheless, for both approaches, the input parameter as hydrogen number density of individual cloud, $n_{\rm H}$, is given only at the cloud surface. This is because the radiation pressure compresses the cloud and the density gradient across the photoionized gas occurs naturally, which is self consistently computed by {\sc cloudy} code. The exact comparison of CP and CD models for the purpose of ILR is given in Sec.~\ref{sec:cpcd}.

For  the radial distribution of clouds, the density number assumed  at each cloud surface changes with distance from SMBH according to a given radial profile of the density at cloud surface $n_{\rm H}(r)$. 
The exact radial density profiles used in our computations are given in 
Sec.~\ref{sec:powlad_dens} and~\ref{sec:real_dens_profile} below.

The ionized clouds are distributed from $r = 10^{-2}$ up to $r =10^{3}$ pc. 
For the given cloud location and for the given density at the cloud surface, the ionization 
parameter $U$ is computed by the {\sc cloudy} code using the well known expression 
\citep{Osterbrock2006}:
\begin{equation}
\label{eq:U}
U=\frac{Q_{\rm H}}{4\pi r^{2}n_{\rm H}~ c}
\end{equation}
where $Q_{\rm H}$ is the number of hydrogen ionising photons in the incident radiation 
field and $c$ is the velocity of light. 

In all our models of radially distributed clouds, we assume that each cloud is illuminated by 
the same shape of SED. We consider four shapes of 
SED adopted from recent multi-wavelength observations of: Sy1.5 - Mrk 509 \citep{Kaastra2011}, Sy1 - NGC 5548 \citep{Mehdipour2015},
NLSy1 galaxy PMN J0948+0022 \citep{Dammando2015},  and LINER - NGC 1097 \citep{Neemen2014}
as displayed in Fig.~\ref{fig:sed_all}.
In order to see spectral shape of different type of AGN,
we normalized all SEDs to $L_{\rm bol} = 10^{45}$ erg s$^{-1}$ in this figure. This value is used in all calculations presented in Sec.~\ref{sec:powlad_dens}, while in 
Sec.~\ref{sec:real_dens_profile} luminosities are taken directly from integrations of 
observations.

All final conclusions of our paper are based on analysis the  modelled line emissivity profiles  
given as an output of photoionization calculation. In the case of each source 
we present radial emissivity profiles  for the most 
observed line transitions in the AGN spectra: H${\beta}$~${\lambda}$4861.36~\AA,
He~II~${\lambda}$1640.00~\AA, Mg~II ${\lambda}$2798.0~\AA, C~III] ${\lambda}$1909.00~\AA, [O~III] 
${\lambda}$5006.84~\AA, Fe~II~${\lambda}$(4434-4684)~\AA, and C~IV~${\lambda}$1549.00~\AA.

\section{Power law density pofile} 
\label{sec:powlad_dens}

\begin{figure}
 \hspace{-4mm}
\includegraphics[width=0.47\textwidth]{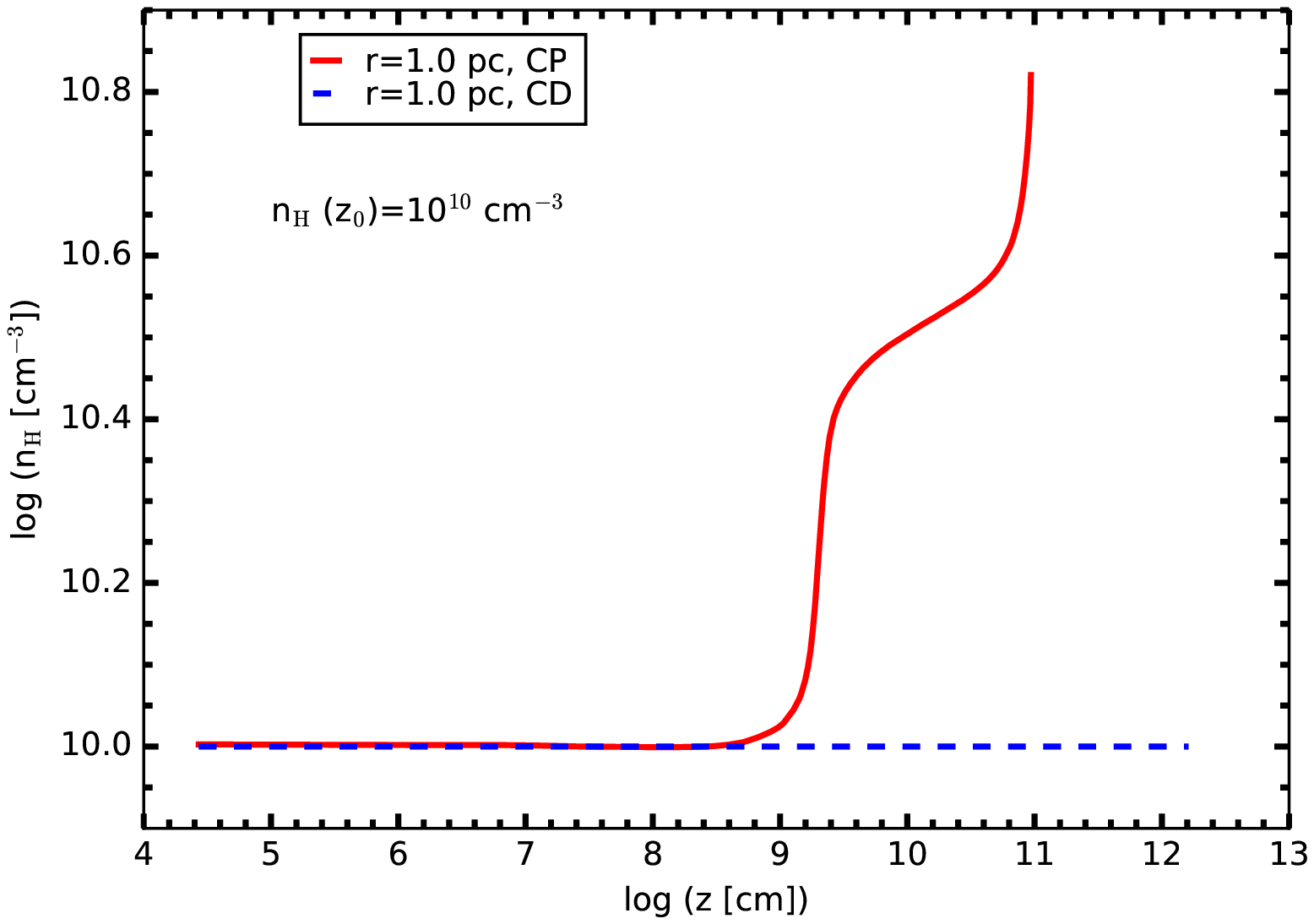}
\includegraphics[width=0.47\textwidth]{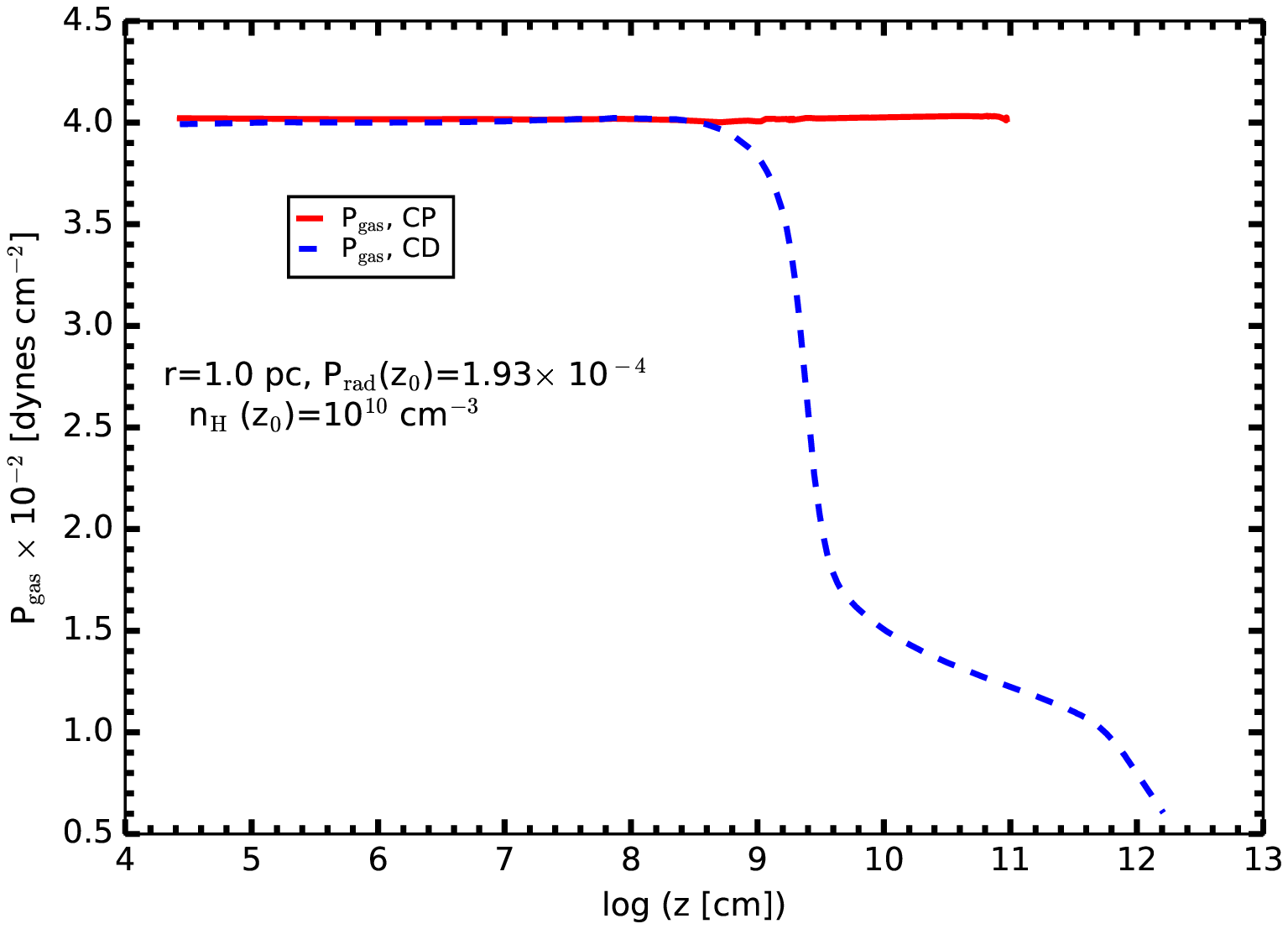}
\caption{\small  The comparison of constant density (CD) and constant pressure (CP) model for a
  sinlge cloud with the density: $n_{\rm H} = 10^{10}$~cm$^{-3}$ at the illuminated
  cloud surface. The Mrk~509 SED is used in both simulations. The density and gas pressure
 stratifications are shown in upper and lower panel, respectively.} 
\label{fig:cpcd10}
\end{figure}

\begin{figure}
 \hspace{-4mm}
\includegraphics[width=0.47\textwidth]{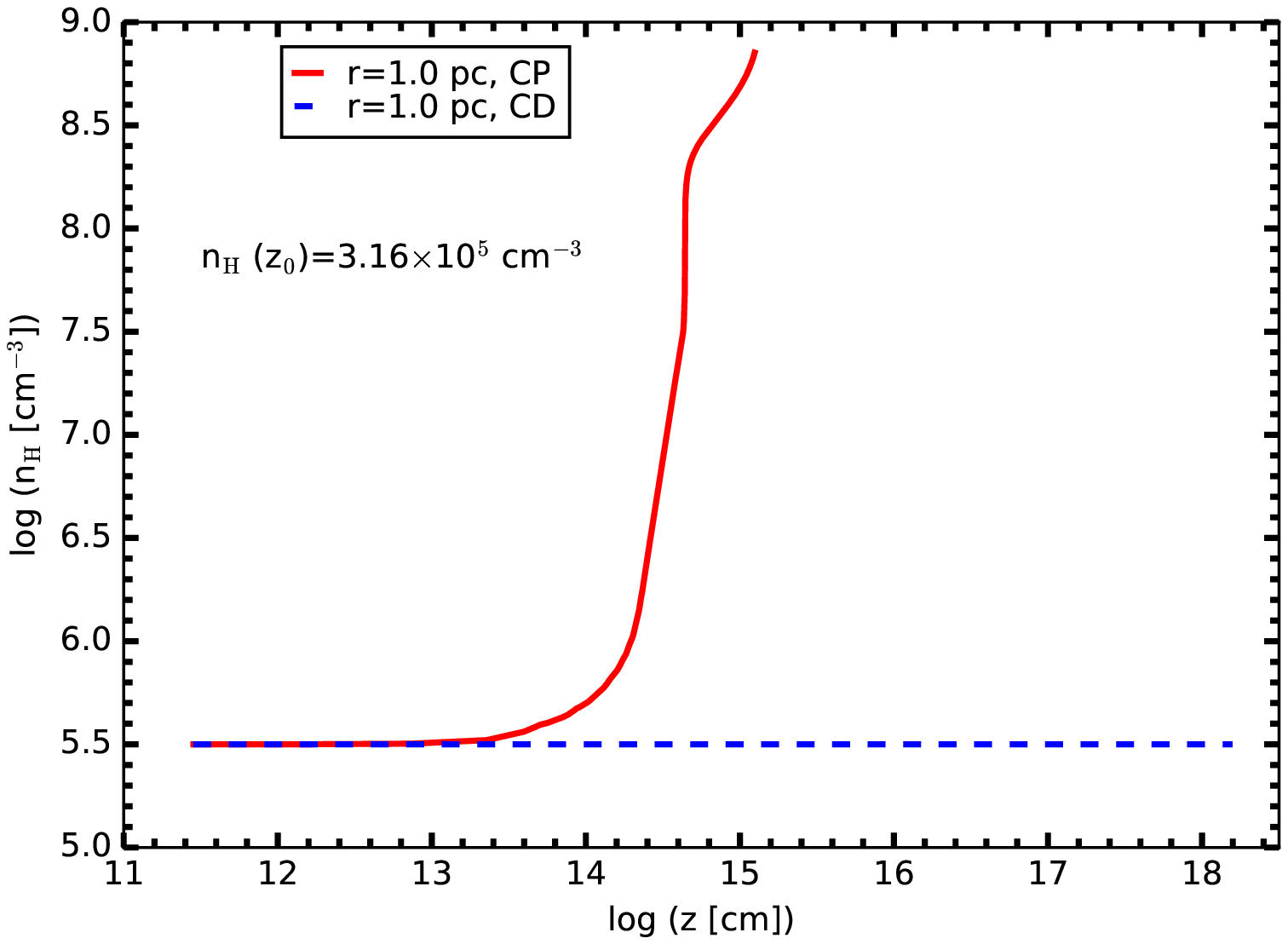}
\includegraphics[width=0.47\textwidth]{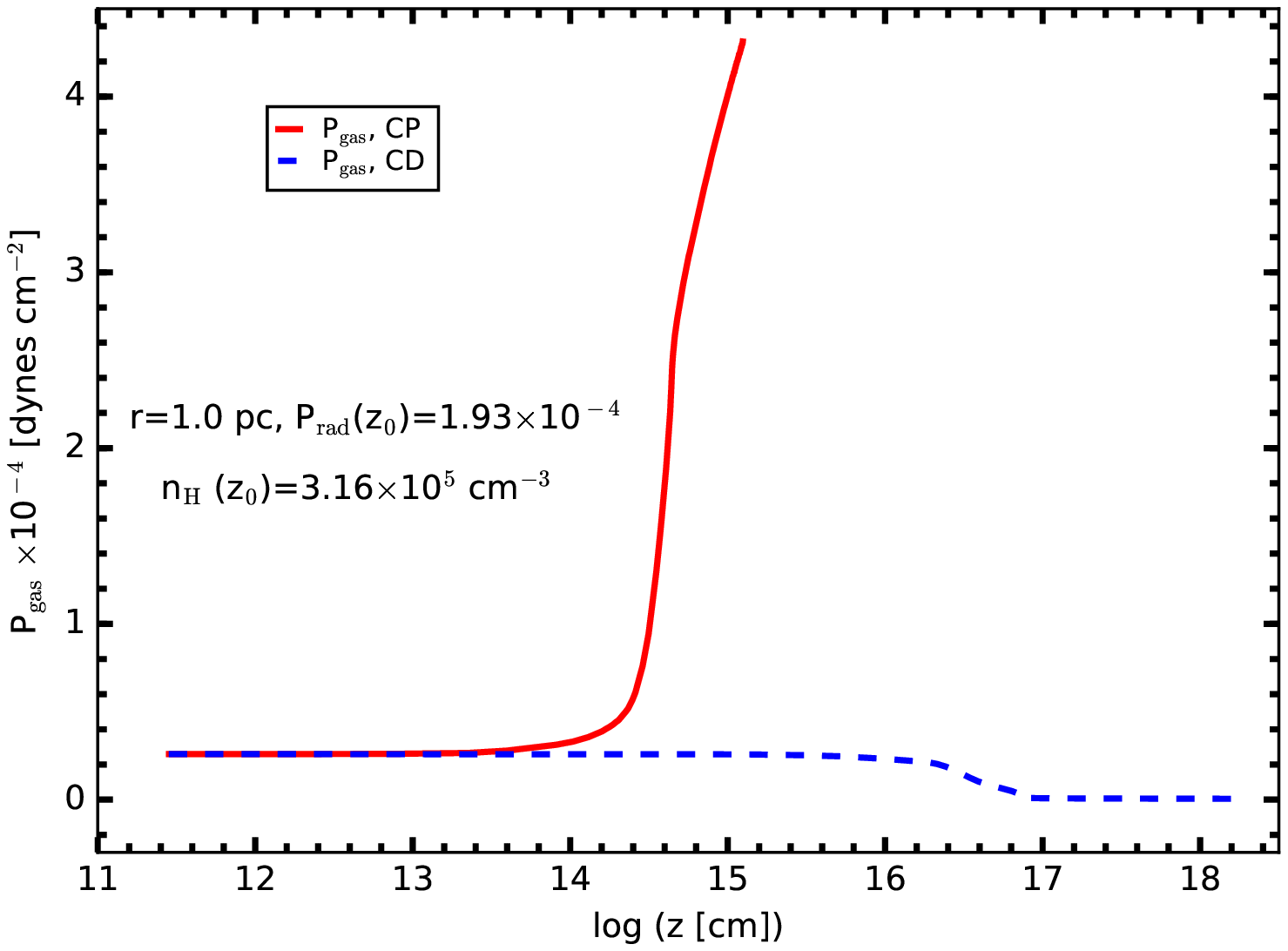}
\caption{\small The same as in Fig.~\ref{fig:cpcd10}, but for low density cloud with:
 $n_{\rm H} = 3.16 \times 10^5$~cm$^{-3}$ at the illuminated
  cloud surface.} 
\label{fig:cpcd5}
\end{figure}

In order to investigate the influence of the density profile on the  
emission line luminosity versus radius, in this section we assume that the density at the cloud surface  
decreases with distance from the SMBH as: 
\begin{equation}
n_{\rm H} = 10^{11.5}(r/R_{\rm d})^{-\beta}
\label{eq:denprof}
\end{equation}
where $\beta$ is the power law density slope. The value of the density at the cloud surface
at the sublimation radius, $R_{\rm d}$, (i.e. density normalization) is adopted after AD16. This is because, AD16 have shown that only for such high density value ILR can exist in the framework of this model. For all models computed in this section 
we adopt the same sublimation radius: $R_{\rm d} =0.1$~pc, following NL93 and AD16. 

 For the density profile given by Eq.~\ref{eq:denprof}, the resulting ionization parameter $U$
depends on the cloud location and on the amount of ionizing photons (Eq.~\ref{eq:U}). 
Assuming the same bolometric luminosity: $L_{\rm bol} = 10^{45}$ erg s$^{-1}$, in case of four sources, we obtain following scaling laws of the ionization parameter with distance in pc:
\begin{equation}
U_{\rm Mrk509}=6.72\times10^{-6} \,(r/R_{\rm d})^{\beta}\,r^{-2}
\end{equation}
\begin{equation}
U_{\rm NGC5548}=3.11\times10^{-6} \,(r/R_{\rm d})^{\beta}\,r^{-2}
\end{equation}
\begin{equation}
U_{\rm NGC1097}=1.13\times10^{-6} \,(r/R_{\rm d})^{\beta}\,r^{-2}
\end{equation}
\begin{equation}
U_{\rm PMNJ0948}=1.08\times10^{-6} \,(r/R_{\rm d})^{\beta}\,r^{-2}
\end{equation}

For the purpose of this paper, we consider three
values of $\beta = 0.5, 1.5, 2.5$, which are taken as an arbitrary numbers to 
relate density as different powers  of distance. We note that $\beta =1.5$ converges to the profile used by NL93 and AD16.

\subsection{CP versus CD for a single cloud}
\label{sec:cpcd}
We consider a single cloud in the spherically symmetric gravitational field, located at radial distance $r$ from the SMBH. 
We assume that locally the cloud thicknes is negligible in comparison to 
the distance which is equivalent to locally 
plane parallel approximation of the cloud geometry. The cloud is illuminated by 
radiation flux $F_0$  at the illuminated face $z_0$. The condition of hydrostatic equilibrium, 
for black hole mass $M_{\rm BH}$, with optical depth as a variable $ d\tau=\kappa \rho dz$, is: 
\begin{equation}
{dP_{\rm gas} \over d \tau} = - {1 \over \kappa} \left({GM_{\rm BH} \over r^2}  - \Omega^2  r \right) - {dP_{\rm rad} \over d \tau}
\end{equation}
where as usual $\rho$ is volume density, $\kappa$ - mean opacity coefficient, $G$ - gravitational constant, 
$P_{\rm gas}$ and $P_{\rm rad}$  - gas and radiation pressure, and $\Omega$ is  the gas angular velocity. Expressing  the radiation pressure gradient as a first order solution of radiative transfer equtation: 
$dP_{\rm rad} /d \tau = - (F_0/c)\,e^{-\tau}$, and integrating of hydrostatic balace 
from 0 to $\tau$, we obtain:
\begin{eqnarray}
P_{\rm gas}(\tau)  &= & P_{\rm gas}(0) - {1 \over \kappa} \left( {GM_{\rm BH} \over r^2} - 
\Omega^2  r \right) \, \tau  \nonumber \\
 & - & {F_0 \over c} \, e^{-\tau} + {F_0 \over c}
\end{eqnarray}
where $c$ means velocity of light.

\begin{figure*}
 \hspace{-4mm}
\includegraphics[width=1.1\textwidth]{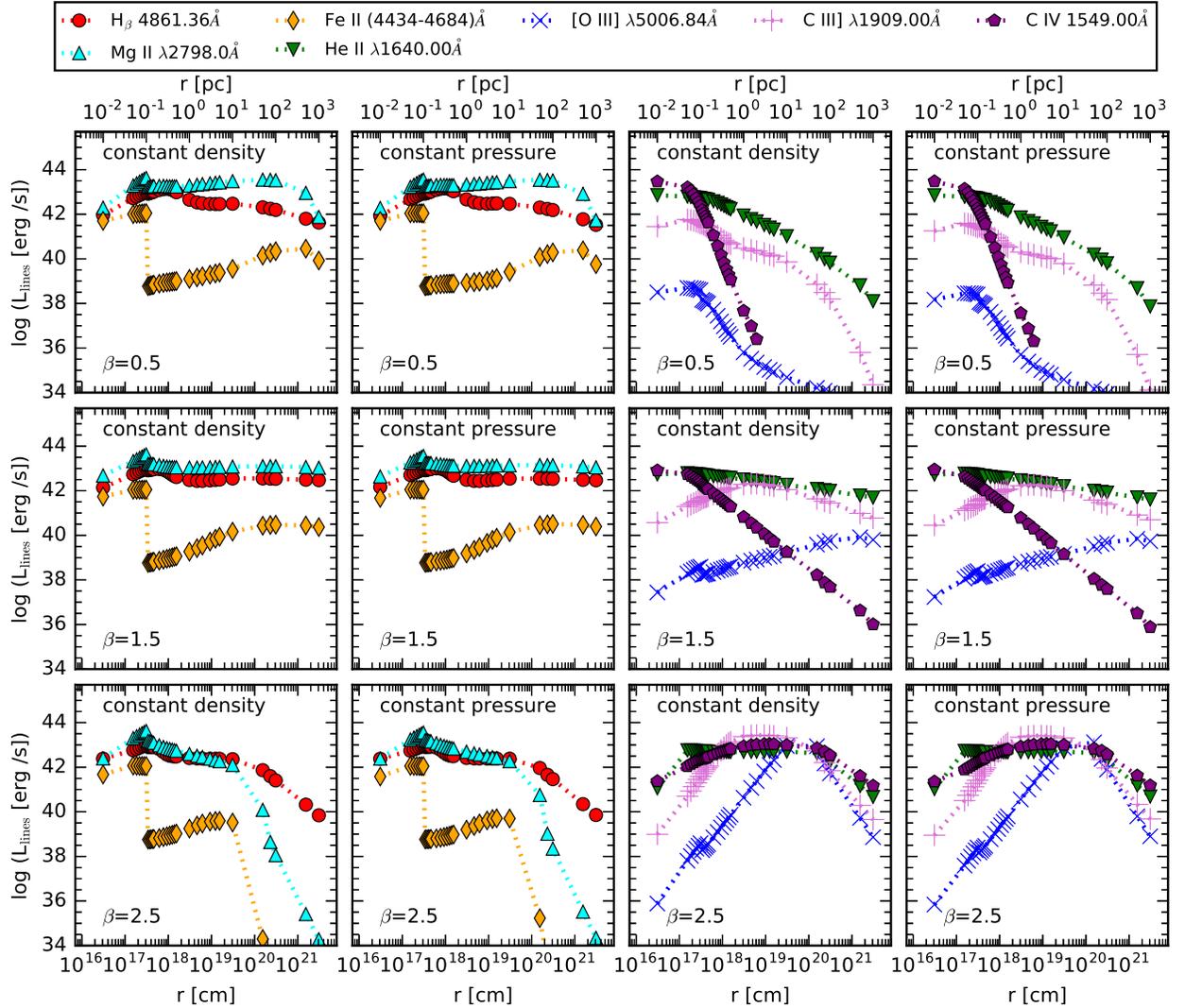}
\caption{\small Line luminosity versus radius for Mrk~509, Sy~1.5 SED.  
Left two column panels represent LIL: H$\beta$ Mg~II and Fe~II lines for constant density and constant pressure case respectively, while right two columns show HIL: He~II, C~III], [O~III], and C~IV lines again for 
constant density and constant pressure respectively. Three row panels show cases for $\beta = 0.5, 1.5$
and $2.5$ from top to the bottom.} 
\label{fig:line}
\end{figure*}

At the illuminated face of the cloud, the total pressure has some constant initial value: $C_0= P_{\rm gas}(0) + F_0/c$. In addition the gas pressure gradient at the outer cloud surface ($\tau = \tau_{\rm max}$) should be zero, since the cloud is finite and the radiation pressure and centrifugal force should balance the gravitational force there i.e. 
$ (GM_{\rm BH}/r^2 - \Omega^2  r)/ \kappa  = (F_0/c) \, e^{-\tau_{\rm max}}$. 
Adopting those conditions the hydrostatic balance is:
\begin{equation}
P_{\rm gas}(\tau) = C_0 - {F_0 \over c} \, (e^{-\tau} + \tau \, e^{-\tau_{\rm max}} )
\end{equation}
To put limits on the initial conditions of cloud pressure which fullfils the 
hydrostatic balance we express the ratio of gas pressure at two extreme cases: 
for $\tau = 0$ and $\tau = \tau_{\rm max}$:
\begin{equation}
{P_{\rm gas}(\tau_{\rm max}) \over P_{\rm gas}(0)} = 
{C_0 - {F_0 \over c} \, e^{-\tau_{\rm max}} \, (1+\tau_{\rm max}) \over 
{ C_0 - {F_0 \over c} }}
\end{equation}

The above equation does not put any limit on the  initial values of cloud pressure being 
in pressure equilibrium. We can consider special limits. Assuming $\tau_{\rm max} \ll 1$, we 
get: 
\begin{equation}
{P_{\rm gas}(\tau_{\rm max}) \over P_{\rm gas}(0)} = 1; \,\,\,\,\, \Rightarrow  P_{\rm gas} = {\rm const}
\end{equation}
which is well known case of constant gas pressure cloud. It means that for optically thin clouds we don't have strong modification of the gas pressure  by the radiation pressure. Second special case occures for $\tau_{\rm max} \gg 1$, and we get:
\begin{equation}
{P_{\rm gas}(\tau_{\rm max}) \over P_{\rm gas}(0)} =   { C_0 \over C_0-{F_0 \over c}} = 
 1 + {P_{\rm rad}(0) \over P_{\rm gas}(0)} 
\end{equation}
In this case the density gradient inside the cloud depends on the adopted value
of gas to radiation pressure at the illuminated face of cloud.
When $P_{\rm rad}(0) \ll P_{\rm gas}(0) $ again  we get condition of cloud being under constant gas pressure. But 
when $ P_{\rm rad}(0) \geq P_{\rm gas}(0)$, the compresion of cloud by radiation pressure is
always present and increases with increaseing value of this ratio. 
Thus the requirement that radiation pressure should be much larger from the gas pressure, made by \citet{baskin2014} for RPC model, is only the special case among solutions for 
constant pressure cloud, where the compression is the strongest. Physically, it is  
always the case of warm absorbers in AGN modelled by \citet{rozanska2008}.
Nevertheless, many other solutions of clouds being under constant total pressure 
are possible from small compression equivalent to constant gas pressure model 
up to strong RPC considered by \citet{baskin2014}  in case of BLR.
 
Physical conditions considered in this paper, needed to give strong emission at ILR,
put us into the limit of $P_{\rm rad}(0) < P_{\rm gas}(0)$ by roughly two orders of magnitude
depending on the cloud location. 
However, constant gas pressure even in this case is not equivalent to constant density model,
since even constant gas pressure can imply large density and temperature gradient.
To illustrate the difference between CP and CD clouds within the framework of our model 
we present in Figs.~\ref{fig:cpcd10} the structure of a single cloud as a result of photoionization
calculations with {\sc cloudy} code. Our cloud is located at 1 pc from SMBH and illuminated by SED
of Mrk~509. The assumed density on the cloud surface is of the order of
$n_{\rm H}=10^{10}$~cm$^{-3}$. The difference between assumption of constant density (dashed blue line)
and constant pressure (solid red line) is noticable. The denisity structure 
for CP cloud  is not constant, even when compression is very weak, since radiation pressure is 
gradually absorbed with cloud optical depth. 

Lower gas pressure at illuminated cloud surface can relatively increase the compression
by radiation pressure. Physically we can achieve this condition for lower density. In
Fig.~\ref{fig:cpcd5} we 
present the same single cloud comparison for density of the order of $10^5$~cm$^{-3}$.
For such case we are in the limit where $P_{\rm rad}(0) \approx P_{\rm gas}(0) $ and
compression is clearly visible as a density and gas pressure rise up  with cloud thickness. 
Nevertheless, recently we have shown that ILR can exist only if density is high \citep{Adhikari2016}, which means our CP clouds are not too strongly compressed, and 
therefore not so different from CD model.  

The requirement of high density cloud at the sublimation radius sets the 
sound-crossing timescale to be two orders of magnitue smaller than the dynamical timescale,
which is expected to be a few years at $R_{\rm d} =0.1$~pc. Cooling/heating 
timescale for such dense gas is even several orders of magnitude shorter 
than sound-crossing timescale, therefore we assume that clouds are both 
in thermal and hydrostatic equillibrum. With such assumption, even dense clouds 
can survive for at least a fraction of the local Keplerian period without being destroyed.

\subsection{CP versus CD for full model}

With above considerations, we derived the line luminosities for the 
major emission lines, which we present in Fig.~\ref{fig:line} for the case of Sy1.5 galaxy Mrk~509 spectral shape. For better visibility we draw emission from LIL:
H${\beta}$, Mg~II, and Fe~II in the left two panel columns, while other HIL as: He~II, C~III], [O~III], and C~IV are show in right two panel columns. 

\begin{figure}
 \begin{center}
\includegraphics[scale=0.73]{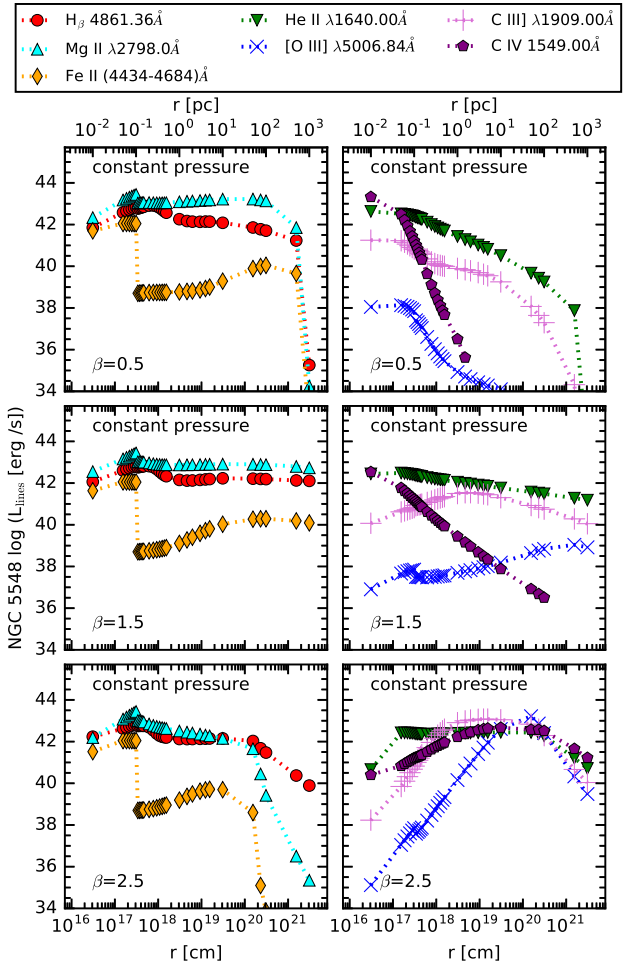}
\end{center}
\caption{\small Line luminosity versus radius for NGC~5548, Sy~1 SED.
Left column panel represents LIL: H$\beta$ Mg~II and Fe~II,  
while right column shows HIL: He~II, C~III], [O~III], and C~IV, for 
constant pressure model. Three row panels show cases for $\beta = 0.5, 1.5$
and $2.5$ from top to the bottom.} 
\label{fig:line1}
\end{figure}

\begin{figure}
 \begin{center}
\includegraphics[scale=0.73]{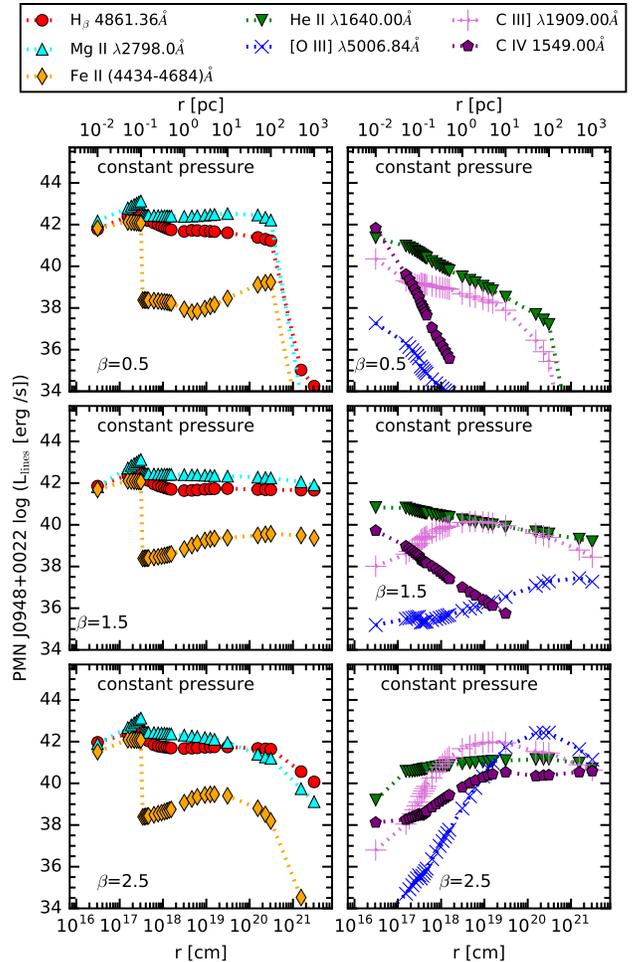}
\end{center}
\caption{\small The same as in Fig.~\ref{fig:line1}, but for PMN~J0948+0022, NLSy1 SED}
\label{fig:line2}
\end{figure}

\begin{figure}
 \begin{center}
\includegraphics[scale=0.73]{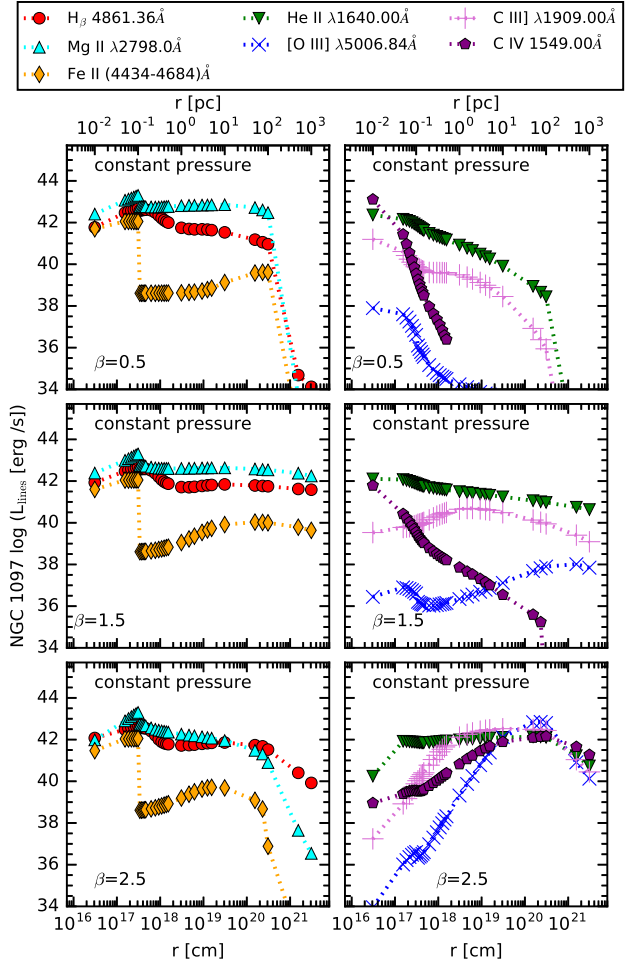}
\end{center}
\caption{\small The same as in Fig.~\ref{fig:line1}, but for NGC~1097, LINER SED.}
\label{fig:line3}
\end{figure}

Each pair of panel columns in Fig.~\ref{fig:line} represents the comparison 
between the model which assumes that each cloud is computed under
constant density, to the model which assumes  CP clouds. We can easily see that 
emission line luminosities do not differ when more physical model of CP is used. 
Profiles of all emission lines are practically the same, when we compare left and
right column of 
both pairs of column panels. We demonstrate here that for such colder clouds as in BLR and NLR,
the compression by radiation pressure is not that important as in warm absorbers studied
by \citet{rozanska2006,Adhikari2015}. 
 As shown in previous subsection, we may expect some differences for lower density 
normalization at a sublimation radius. But for lower density we are not able to produce 
visible ILR region, which is a purpose of this paper.
The above conclusion is valid for all four spectral shapes used in this paper, therefore for other 
AGN types we present only CP models of line emissivity in Figs.~\ref{fig:line1}, \ref{fig:line2}, and \ref{fig:line3}.

\subsection{The density power-law slope}
\label{sec:slope}

Only for $\beta=1.5$, the overall line emissivity profiles stay flat along the radius 
in case of four AGN types, with exception of Fe~II and C~IV lines.  
The density profile with such power law slope changes from $\sim 10^{12}$~cm$^{-3}$ in BLR to 
$\sim 10^6$~cm$^{-3}$ in NLR with the later value favorable by narrow lines. 

The situation changes when  $\beta=0.5$, and density still is high $\sim 10^{9.5}$~cm$^{-3}$
in NLR. All LIL as H$_\beta$, Mg~II and Fe~II, presented always on left column panels,
display sudden drop in emissivity in the NLR range $r \gtrsim 100$ pc, caused by the low value of ionization parameter (due to high density). 
The exception is Mrk~509 (Fig.~\ref{fig:line}, because there are many of UV photons in its spectral shape
(Fig.~\ref{fig:sed_all} dashed-red line). 
The high value of those photons still keeps the ionization parameter high enough to produce 
strong LIL. On the other hand HIL, presented always on right column panels
decrease monotonically with distance.   

On the second extrema, when $\beta=2.5$, the density is very low $\sim 10^{1.5}$~cm$^{-3}$ in NLR. 
This provides to the visible drop of LIL in the  NLR range $r \gtrsim 100$ pc, caused by too low density. It happens in all types of AGN. Nevertheless, the emission of HIL increases with the distance from SMBH up to the point about $r \sim 10$ pc due to relatively high ionization parameter. Further away from the center, such emission becomes flat or decreases
depending on the value of ionization parameter. 
   
In general results do not depend much on SED shape and in all cases ILR is visible with exception 
of Fe~II permitted and [O~III] forbidden lines. Forbidden lines are effectively produced in 
low density environment and for many cases presented here their emissivity is too low.      

\subsection{Dust sensitive Fe~II line}
\label{sec:fe2}
 
In all cases, the Fe~II line is the only line which shows strong emissivity drop by several orders 
of magnitude at the sublimation radius. Such behavior predicts the lack of intermediate component 
for this line. Therefore, based on the results of 
our simulations, the Fe~II line is sensitive to the presence of dust,  and it is not ILR indicator. 

\begin{figure}
 \hspace{-4mm}
\includegraphics[width=0.53\textwidth]{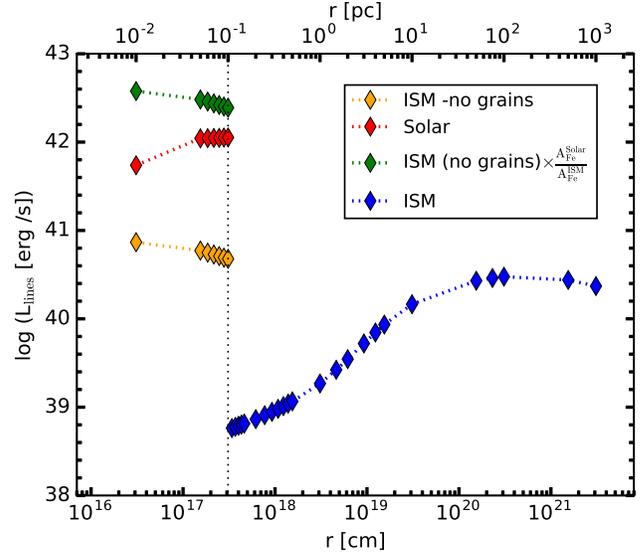}
\caption{\small The comparison of Fe~II line luminosity for two models of clouds 
illuminated by Mrk~509 SED with different abundances of dustless clouds. The ISM composition 
with grains is used for dusty clouds located further than $R_{\rm d}$, marked by vertical dotted line. For clouds located closer to SMBH than $R_{\rm d}$ we plotted Fe~II line luminosity for Solar iron abundance by red diamonds and for ISM without grains abundance 
by orange diamonds. Green diamonds mark the ISM model without grains multiplied by the Fe abundance ratio of those two models.} 
\label{fig:abun}
\end{figure}

In our model, strong Fe~II emissivity drop may be caused by two effects. The first one is the presence of dust discussed above, but the second effect can be the change of abundances in the gas phase when passing 
sublimation radius. The assumed solar composition for $r<R_{\rm d}$ has two orders of magnitude
higher iron abundance than ISM composition for $r>R_{\rm d}$. All other elements display the same magnitude abundances when changing from solar to ISM composition. The change of abundances mimics the 
depletion of metals due to dust sublimation as it was already assumed by NL93 and AD16.
In order to check what really causes the strong drop 
of Fe~II emissivity profile at $R_{\rm d}$, we made a test with different abundances in dustless
clouds located at BLR. Results are presented in Fig.~\ref{fig:abun}, where we compare two models. 
In the first model, we assume ISM chemical composition with no grains for BLR clouds, while 
in the second model, the same clouds have typical solar abundances. In both models ISM composition 
with grains is used for dusty clouds located further than $R_{\rm d}$. It is clearly seen that 
the dust present is responsible for  the Fe~II emissivity drop by about two orders of magnitude, 
while the change in iron abundance enhances this effect by one order of magnitude.

\section{Disk-like density profiles}
\label{sec:real_dens_profile}

Since many years, it was postulated that at different radii we should have outflows from disk atmosphere 
in AGN \citep[i.e.][]{elvis2004}.
In this section we consider disk-like density profile, $n_{\rm H}(r)$, which is expected where clouds 
are formed from outflowing gas above the accretion disk atmosphere. 
 We do not specify the mechanism which provides to the formation of 
clouds, we only assume they do exist and they should have the same density as upper disk atmosphere. 
For determination of cloud disk-like density profile, we 
have to specify  disk parameters, as black hole mass and accretion rate for each type of AGN. 
Table~\ref{tab:param} displays all important values used in our further computations. 

By adopting black hole masses and accretion rates, we simulate the vertical accretion disk 
structure assuming standard \citet{SS73} disk and transfer of radiation by diffusion approximation with gray gas opacities as described by \citet{rozanska99}.  
Furthermore, we employ the cloud radial density profile  by the requirement that 
the adopted gas density at the illuminated cloud surface equals to the disk atmosphere density where $\tau=2/3$, i.e. the atmosphere is still optically thick.
When this comparison is done at each distance from the black hole, we obtain disk-like cloud
density profile.  

\begin{figure}
\includegraphics[width=0.51\textwidth]{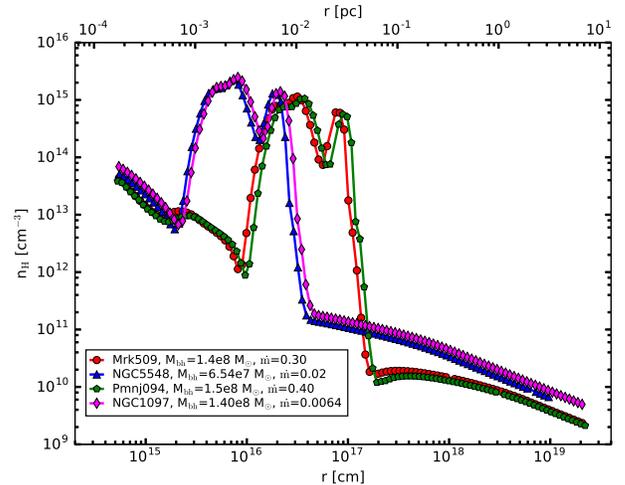}
\caption{\small Radial density profiles as expected from the 
upper zones of accretion disk atmospheres for all considered sources: 
Parameters used in computations of the corresponding density profiles are also shown
in the legend box.} 
\label{den:disk}
\end{figure}

\begin{figure}
\includegraphics[width=0.51\textwidth]{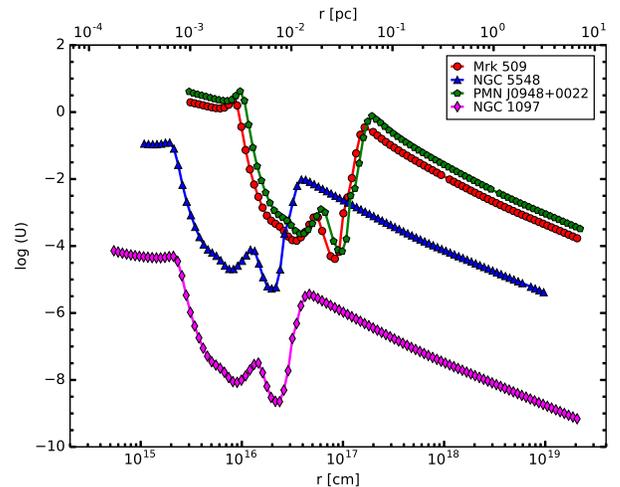}
\caption{\small Radial ionization parameter $U$ profiles at the cloud surface computed
by Eq.~\ref{eq:U} for disk-like density profiles as shown in Fig.~\ref{den:disk}. } 
\label{den:ion}
\end{figure}

The disk-like density profiles for all sources are presented at Fig.~\ref{den:disk}.
The characteristic feature of such radial density profiles posses strong density rise 
up to $10^{15}$ cm$^{-3}$, located around the position of BLR, $r\sim 10^{-2}$ pc. 
This is caused by strong opacity hump in an accretion disk atmosphere.
Outside density hump its values decline from about $10^{13}$ cm$^{-3}$ at the distance of $10^{15}$~cm 
from black hole, to about a few $10^{9}$ cm$^{-3}$ at further distances of $10^{19}$ cm.
The corresponding ionization parameters for each disk-like density profiles are presented 
in Fig.~\ref{den:ion}. One can see noticable difference between ionization degree, which 
is four orders of magnitude lower for LINER than Sy1 and NLSy1 case. In addition, the density hump is directly reflected in the ionization drop in all sources.

In this section, the source luminosity, $L$, given in third column of Table~\ref{tab:param} 
is obtained by the integration of flux between the energy 
range from 1 to 10$^5$ eV. Furthermore, the dust sublimation radius, $R_{\rm d}$ for each 
source luminosity is computed 
using the following formula given by \citet{Nenkova2008}:
\begin{equation}
\label{eq:rdust}
R_{\rm d} = 0.4 \sqrt{L/10^{45}} ~~~~~[{\rm pc}].
\end{equation}
This formula simply indicates the radius at which, for a given luminosity, the gas temperature reaches the value of 1400~K. Below this temperature dust can survive 
as a substantial gas component. The sublimation radius corresponding to each type of AGN is given in 6th column of Table~\ref{tab:param}.
It is clear that the luminosity influences the position of sublimation radius, 
which we fully take into account in this section.
In addition, the normalization of $N_{\rm H}$ to 10$^{23.4}$ cm$^{-3}$ differs 
for each source, since it is set at the position of sublimation radius. 

\begin{table}
\caption{\small Parameters used in computations of disk-like density profile and
the position of $R_{\rm d}$. The first and second column lists the name of AGN and its type. 
The integrated luminosity is given in third column. The black hole masses in $10^7$~M$_{\odot}$,
and accretion rates in units of Eddington accretion rate, follow in fourth and fifth column respectively. 
Both values are taken from literatures listed below. 
The derived dust sublimation radius in $10^{17}$ cm is given in column six.}
\label{tab:param}
\begin{tabular}{lllccl}
\hline 
Name & AGN & $L_{(1-10^5){\rm eV}}$ & $M^{\rm BH}_{7}$  & ${\dot m}$ & $R^{\rm d}_{17}$ \\
 & type & erg s$^{-1}$ & M$_{\odot}$ & $\dot {\rm M}_{\rm Edd}$& cm \\
\hline
Mrk~509 &  Sy1.5 & 6.62$\times 10^{45}$ & 14$^a$  & 0.30$^{b}$ & 31.6  \\
NGC 5548 &  Sy1 &1.28$\times 10^{44}$   & 6.54$^c$ & 0.02$^d$ & 4.41 \\
PMN J0948 & NLSy1&2.28$\times 10^{46}$ & 15.4$^e$ & 0.40$^f$ & 58.9 \\
NGC~1097 & LINER &9.62$\times 10^{40}$ & 14$^g$ & 0.0064$^h$ & 0.12 \\
\hline 
\end{tabular}
a -- \citet{Mehdipour2011},
b -- \citet{Boissay2014},
c -- \citet{Bentz2007},
d -- \citet{Crenshaw2009,Ho2014},
e -- \citet{Foschini2011},
f -- \citet{Abdo2009},
g -- \citet{Onishi2015},
h -- \citet{Neemen2014}.

\end{table}

The position of sublimation radius strongly depends on the detailed dust composition which is still
under discussion \citep{gaskell2017,xie2017}. The dust is most likely the mixture of amorphous carbon
\citep{czerny2004}, silicate \citep{lyu2014}
and graphite grains \citep{baskin2018}, while the sublimation radius derived by \citet{Nenkova2008}
corresponds to the temperature of sublimation of silicate grains only \citep{laor1993}. Graphite
grains sublimate at larger temperatrue up to $\sim 2000$~K  \citep{laor1993,baskin2018},
nevertheless AGN extinction curves do not show the 
2175\AA~carbon feature \citep{maiolino2001} which makes the dust in the circumnulear region of
AGNs being dfferent from Galactic ISM.  Nevertheless, to show how our results do depend on
the dust sublimation radius, below we present our 
model computed for two sources with sublimation radius about 10 times lower: 
$R_{\rm d} = 0.06 \sqrt{L/10^{45}}$~pc, which corresponds to graphite sublimation 
temperature \citep{laor1993}.

We assume that emitting clouds directly emerge from the disk's 
atmosphere and preserve its density along the whole range of radii. This is reasonable 
assumption at least for low ionization part of the BLR as it may develop as a failed 
wind embedded in the disk's atmosphere \citep{czerny2011}.  On the other hand, this
assumption is the same as the model of BLR being a part of accretion disk 
atmosphere \citep{baskin2018}.
Our approach is not in contradiction with line emitting medium geometry similar to bowl on top of the accretion disk \citep{Gaskell2009,Goad2012}.

\begin{figure}[h]
\hspace{-1mm}
\includegraphics[width=0.45\textwidth]{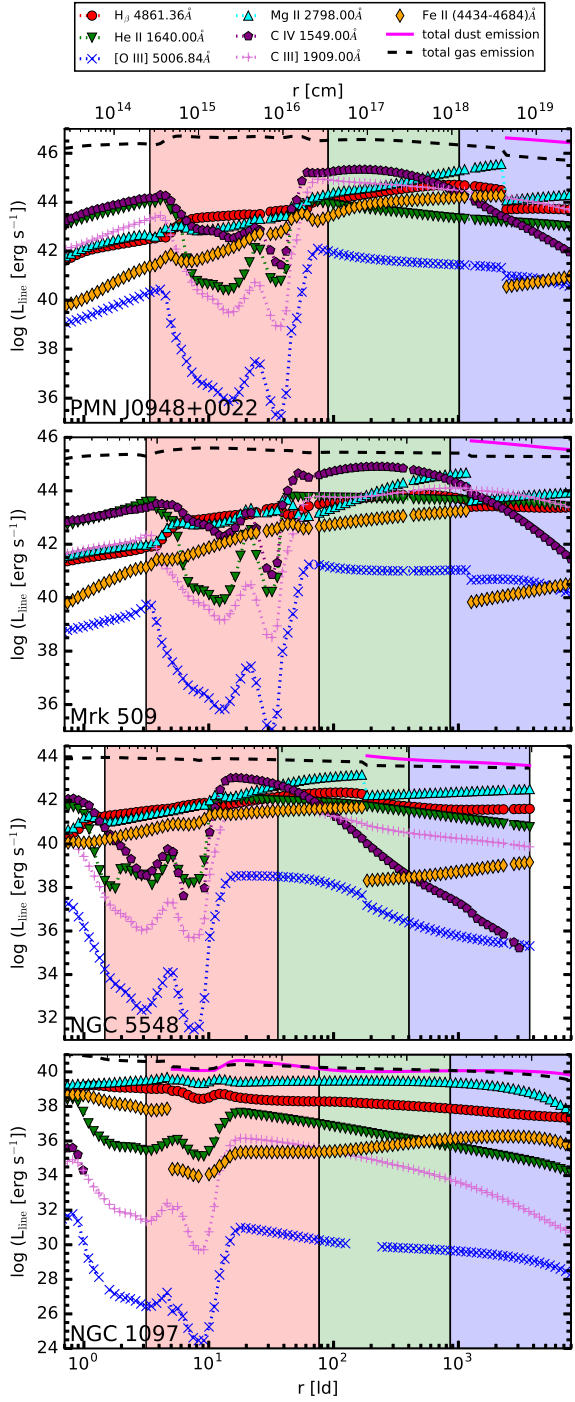}
\caption{\small LIL and HIL luminosities versus radius (in light days) obtained for disk-like cloud density 
profiles. Each panel shows one type of AGN in the way that the source luminosity (given in Table~\ref{tab:param}) decreases from the top to the bottom panel. 
Shaded areas mark the position of BLR, ILR and NLR from the left to right respectively, based on the addopted range of Keplerian velocities (see text for details).  
}
\label{fig:rev}
\end{figure}

\begin{figure}[h]
\hspace{-1mm}
\includegraphics[width=0.45\textwidth]{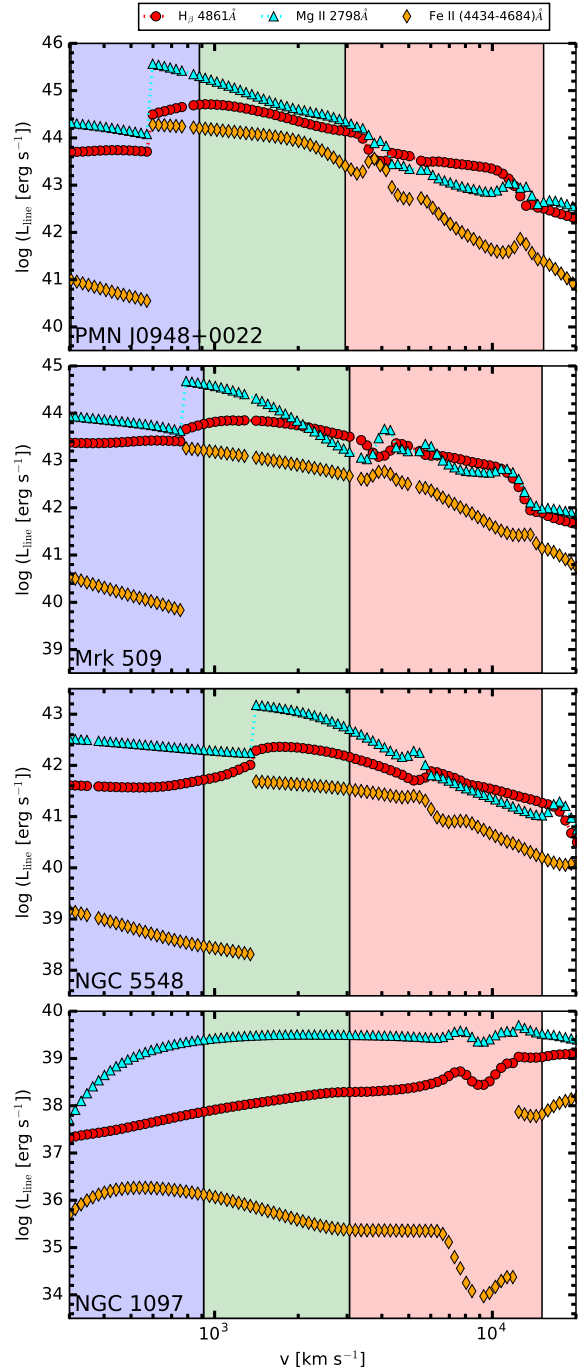}
\caption{\small LIL luminosities versus Keplerian velocity obtained for disk-like cloud density 
profiles. Each panel shows one type of AGN in the way that the source luminosity (given in Table~\ref{tab:param}) decreases from the top to bottom. Shaded areas mark the position of BLR, ILR and 
NLR from the right to left respectively,  based on the addopted range of Keplerian velocities (see text for details).
}
\label{fig:kep}
\end{figure}

Resulting line emissivity profiles for four sources, are presented in Fig.~\ref{fig:rev}
-- HIL and LIL, and Fig.~\ref{fig:kep} -- only LIL. In addition, for each source we present total dust and total gas emission by magenta solid and black dashed lines respectively. Shaded areas on the figures mark the position of BLR, ILR and NLR 
which depends on the AGN type. Assuming that all emitting gas is dominated by Keplerian motion, 
BLR marked with pink shadow spans between 15000~km~s$^{-1}$ down to 3000~km~s$^{-1}$. 
ILR marked with green shadow spans between 3000~km~s$^{-1}$ and 900~km~s$^{-1}$, while NLR 
marked with violet shadow spans between 900~km~s$^{-1}$ down to 300~km~s$^{-1}$. 

In all AGN types, the emissivity profiles of LIL are insensitive to the 
density hump in disk-like cloud density profile. On the other hand, HIL display strong
luminosity drop which reflects the density enhancement in the cloud radial profile. 
Such HIL luminosity drop is usually situated in BLR indicating the division of BLR on two types
of low and high ionization as previously suggested by \citet{collin88,Collin2006,czerny2011}.

In case of PMN~J0948 and Mrk~509, luminosity is high enough to push sublimation radius above 0.1~pc 
making part of the ILR free from dust. This allows for appearance of emissivity maximum within 
ILR below sublimation radius (three upper panels of Fig.~\ref{fig:rev}). 
Thus, in Seyferts, our model predicts dominating intermediate component in LIL: 
H$\beta$,  Mg~II, and Fe~II. Their line luminosities rise monotonically up to 
sublimation radius.  Nevertheless, the emission of HIL: He~II, C~III], [O~III], 
and C~IV is maximal in outer BLR or inner ILR, and decreases with distance. 

In case of LINER, dust sublimation radius appears at the position of HIL emissivity drop, caused
by cloud density hump, because of its low luminosity. The presence of dust in BLR region provides to flat line emissivities all the way to NLR. Therefore, ILR could be present 
in LINER, but it is not as strong as in other types of AGN. Our result confimrs 
the statement made by \citet{Balmaverde2016}, where the authors concluded that 
high density component of inner portion of ILR is visible in Seyferts whereas 
the entirety of ILR emission is visible in LINER. Nevertheless, the density of 
ILR clouds in LINER inferred by those authors is 10$^{4-6}$~cm$^{-3}$, 
less by $\sim$ 5 orders of magnitude than the density of the ILR clouds in our model.
We predict for NGC~1097, the location of ILR at the range of radii 0.07--0.8~pc whereas, the 
distance  of ILR inferred by \citet{Balmaverde2016} is 1--10~pc.

In general, for disk-like density profile, additional lines as Mg~II and slightly H$\beta$, 
appear to be dust sensitive. Their luminosity profiles exhibit rapid decrease when dust appears 
in clouds located relatively far from the center. This does not happen in case of LINER, since 
dusty clouds are still very dense (Fig.~\ref{fig:kep} fourth panel).

The disk-like density cloud distribution model is less general, than arbitrarily 
taken density profile in locally optimal clouds model \citet{Goad2014},
 but it heavily depends on the conditions in disk's upper atmosphere. The results 
are also different. While in our model emissivity of  H$\beta$ line increases with distance by two orders of magnitude for
Seyfert AGN, \citet{Goad2014} has shown decline of this line between 1 and 100 light days. 
The consequence would be that our model predicts stronger importance of ILR component.

\section{Emission line regions}
\label{sec:lines}
Syfert 1 -- NGC~5548 is one of the best studied AGN. It has been regularly monitored for 
almost four decades. 
Thus it is clear that model of this source could be discussed most critically.
We can compare approximate distances of emitting regions from our model with the results of reverberation mapping campaigns. We are aware of, that reverberation mapping depends on the continuum luminosity
measurement \citep{Peterson2004,Bentz2006,Bentz2007,Denney2009}, and data show delays 
measurements in H$\beta$ 
span over 6-30 days depending on the continuum luminosity. 
Therefore, in case of H$\beta$ line, we made analysis for continuum luminosity, $L_{\lambda}(5100$\AA),
derived  from incident SED (Fig.~\ref{fig:sed_all}), and  have checked what radius of the H$\beta$ emission we should expect from observational measurements \citep{Kilerci2015}. For log$L_{\lambda}(5100$\AA)~$\approx 43.25$ used in our model, we have delay in H$\beta$ of the order of 20 days and line width 
FWHM$\approx 4700$~km~s$^{-1}$. Similar results were obtained by \citet{Peterson1991} using ground-based observations made in 1989, where 21 days delay between continuum 
and H$\beta$ was reported. Those observed parameters correspond to the radius at which the H$\beta$ line luminosity  reaches maximum at about $4 \times 10^{17}$~cm in our model (Fig.~\ref{fig:rev} third panel).

\begin{figure}[h]
\hspace{-1mm}
\includegraphics[width=0.45\textwidth]{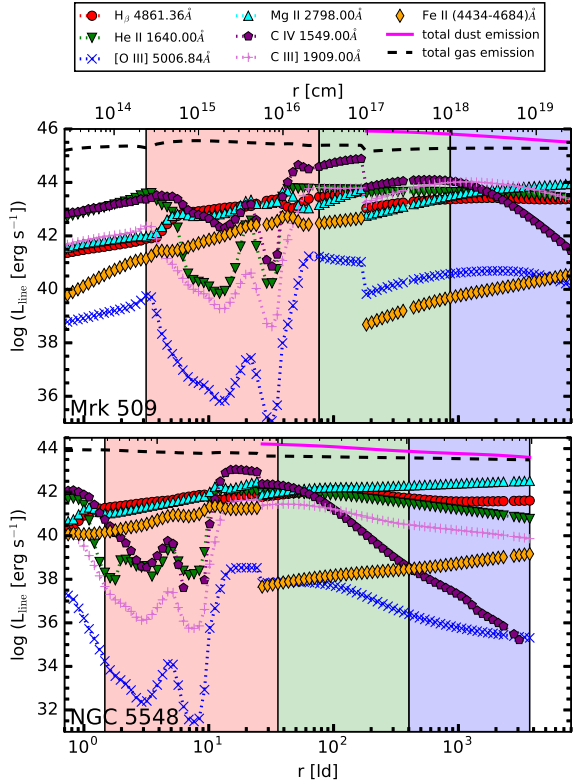}
\caption{\small The same as in Fig.~\ref{fig:rev} for two best observed sources, Mrk~509 (upper panel) and MGC~5548 (lower panel), but for sublimation radius computed from the formula:
$ R_{\rm d} = 0.06 \sqrt{L/10^{45}}$ \citep{laor1993}.  
}
\label{fig:rev2}
\end{figure}

However recent observations of velocity resolved reverberation mapping presented by \citet{Pei2017} show 
new points which do not agree 
with the fitted trend of $L_{\lambda}(5100$\AA) versus H$\beta$ delay \citep{Kilerci2015}. Those new points for log$L_{\lambda}(5100$\AA)~$\gtrsim 43.3$ present H$\beta$ delay of the order of 3 days, so shorter than even for the lowest, log$L_{\lambda}(5100$\AA)~$\approx 42.5$, luminosity state. 
\citet{Lu2016} derived BLR response delay to the source luminosity change to be 2.4~years. 
During this time, BLR may be rebuilt under change of radiation pressure, thus making the comparison 
of our model to the data more difficult.

NGC~5548 emissivity profile of both low ionization lines H$\beta$ and Mg~II is very similar 
in our model (Fig.~\ref{fig:kep} third panel). 
However it is not exactly the case in reverberation mapping observations. Mg~II line is more puzzling 
in this case. \citet{Clavel1991} presented peak-center delays from multimonth IUE campaign done 
in 1989. They have found very broad response in Mg~II line covering 34--72 days. 
In addition, \citet{Cackett2015}, analyzing Mg~II variability have found only weakly correlated broad response to the continuum brightening, with delay response spanning 20--70 days range. Both results may suggest that line luminosity global maximum is located in ILR, which fully agrees with
our model. However, even shallower, global line luminosity 
maximum in H$\beta$ located at the same region of our model is not resolved in the
observations of delay. Mg~II shows more luminous ILR with maximum before face of the torus.
Brighter ILR in magnesium line should be reflected in higher average delays than in hydrogen
Balmer lines and this is the case in reverberation measurements.

For the case of NGC~5548, \citet{Clavel1991} presented light-curves of C~III] and C~IV for 
which the delay covers approximately 26--32 days and 8--16 days respectively. 
This is consistent with the position of global maximum in the emissivities of 
those lines further on and at the outer edge of dense BLR, which fully agrees with our 
model. This is also explained by \citet{Negrete2013}, who inferred emission radius of those high 
ionization lines from the photoionization condition. 
They found optimal emission of C~IV for $n_{\rm H} = 10^{12}$~cm$^{-3}$, log$U \approx -2$ 
and C III], and for $n_{\rm H} = 10^{10}$~cm$^{-3}$, log$U \approx -1.5$ which is 
consistent with our model as seen in Fig.~\ref{den:ion}.
C~III] emissivity in ILR is rather flat and allows noticable emission originating from clouds located at higher radii.

The permitted He~II line is always broad and blended with semi-forbidden O~III], therefore the measurements of delay covering 4-10 days are more difficult to explain by our model. 
Such delay corresponds to the location of emissivity drop of He~II line in the dense BLR 
in NGC~5548 (see third panel of Fig.~\ref{fig:rev}). Our model predicts that
the He~II luminosity local maximum is located on the outer 
edge of the dense BLR for expected $\approx$ 20 days delay.

HST monitoring described by \citet{Rosa2015} reveal C~IV delay of 5 days and He~II delay of
2.5 days. This again corresponds to the emissivity drop of our model.
Those results may demand stronger 
modification of our model. For instance, gasous clouds presented on higher
elevation above the disk where density departure from the atmospheric value may be significant.
Such geometry for broad high ionization line regions were already postulated in the 
literatures \citep[i.e.][]{Collin2006,Decarli2008,Kollatschny2013}.

 To check how the possition of dusty torus (discussed in Sec.~\ref{sec:real_dens_profile}) influences
 radial emissivity profile, in Fig.~\ref{fig:rev2} we present the case of two best studied sources
 computed for 10 times smaller value of sublimation radius taken as
$ R_{\rm d} = 0.06 \sqrt{L/10^{45}}$ \citep{laor1993}. For Mrk~509, \citet{Koshida2014} 
have reported 120-150 days delay of dust phase, depending on the method of derivation.
This value fully agrees with the position of sublimation radius from the formula by 
\citet{laor1993}, presented in upper panel of Fig.~\ref{fig:rev2}. However, this fact would eliminate dominance of 1000 km/s component in Mg~II line and possibly would remove intermediate width line component of global emissivity maximum of H$\beta$. This brings our model closer to the observed line delays.
In case of NGC~5548, \citet{Koshida2014} derived inner torus face radius comparing optical and 
near infrared variability. Their measurements cover range from 60 to 80 
days depending on the NGC~5548 continuum luminosity and method used in 
calculations. This is over 3 times larger than sublimation radius by \citet{laor1993}
(lower panel of Fig.~\ref{fig:rev2}, and 2  times smaller than the one computed by Eq.~\ref{eq:rdust}. 
The lower position of sublimation radius in NGC~5548 influences maximum emissivity of Mg~II shifting it
to the lower radii and decrease contribution from 2000 km/s component.

\citet{Peterson2013} investigated variability of forbidden [O~III] line in NGC~5548. 
They found delay between 10-20 
years ($\sim$2-3 pc) and suggested emitting medium with density $10^5$ cm$^{-3}$. While NLR studied in X-rays, 
as suggested by \citet{Detmers2009}, covers 1-15 pc or more precisely 14 pc as derived by \citet{Whewell2015}. 
In our model densities corresponding to NLR remain high ($\approx 10^9$ cm$^{-3}$). We have high 
emissivities of narrow components in all permitted lines except C~IV. In addition, [O~III] emissivity remains rather low. \citet{Crenshaw1993} reported strong narrow components in all optical/UV permitted lines, especially C~IV. Thus our model is less accurate reproducing NLR, for assumed disk-like density 
profile.

Mrk~509 model is in many aspect similar to the NGC~5548. The most noticeable difference is shift in
the emissivity maximum of C~III] and C~IV toward greater radii. This predicts stronger intermediate 
emission line components from our model.  And this seems to be the case when we look at the observational spectra, presented for 
example by \citet{Negrete2013}. Our model computed for NLSy1--PMN~J0948 shows very similar emissivity 
profile shapes to those of Mrk~509 as those sources have similar BH masses but different SEDs. 
However, line luminosity to the continuum luminosity ratio is lower for NLSy1, thus effectively 
broad components blend with continuum and only contrast of narrow components 
remain sufficient to make line visible.
The LINER case of NGC~1097 is exceptional because the sublimation radius is inside our dense BLR. This fact makes all line luminosity profiles flat up to NLR, which is in agreement with \citet{Balmaverde2016} who emphasized the extended ILR in LINERS up to 10~pc.
In case of flat radial luminosity profile, narrow component has the highest contrast, therefore  it will dominate line profile, which is in agreement with \citet{Gonzales2015}, who pointed out that AGN dominated LINERs are very similar to Seyfert 2 galaxies.

\section{Conclusions}
\label{sec:conc}
We carried out the photoionization simulations of the ionized gas clouds 
in AGNs and studied the effect of varying density profiles on the line 
emission across the radial distance that spans all the way from BLR down
to the NLR. The different density prescriptions are applied in the following ways:
a) we employed the density profile as simple power law used
by NL93 and AD16, and varied its slope. b) We self consistently computed the disk-like density profiles for each AGN by using their observed properties; black hole mass and Eddington ratio.

Using the various density profiles derived, we computed the line luminosities of the 
major emission lines in Sy1.5 Mrk~509, Sy1 NGC~5548, 
NLSy1 PMN~J0948+0022 and LINER NGC~1097 differing by their SEDs. Below
we list final conclusions:
\begin{enumerate}
\item  In case of clouds located at distances considered in this paper, CP and CD cloud models
  reproduce exactly the same  line luminosity profiles in the regime  of lines observed in optical/UV.
It is caused by the fact that our clouds are dense. Lower density clouds do not produce the ILR.
\item The varying slope of the power law density profile does not affect the nature of the ILR. In particular,
the intermediate emission in H$\beta$ is present for all the slopes independent of the SED
shape.
\item For the lower slope of the density profile, forbidden [O~III] and semi-forbidden C~III]
lines are strongly suppressed because of the high density environment. As the slope becomes 
more steeper, i.e. density decreases, these lines are prominent at 
radial distances corresponding to NLR.
\item Fe~II emission line appeared to be most sensitive on the dust presence, since its luminosity 
drops by two orders of magnitude at the sublimation radius (see  Sec.~\ref{sec:powlad_dens} for details).
\item The density drop in the disk-like density profiles causes mild 
enhancement of Mg~II, H$\beta$ and Fe~II lines,  while He~II, C~III] and [O~III] are suppressed at the density drop location. This result is consistent with separation of LIL and HIL clouds in two-component BLR model \citep{collin88}. 
\item The low luminosity of the LINER NGC~1097 shifts the dust sublimation radius 
toward smaller distances from SMBH, which makes the emissivity profiles of all lines flat.
Therefore, intermediate line component can be detectable, but is less prominent than the narrow line component.   
\item The distance inferred from the time delay of H$\beta$, Mg~II, in NGC~5548 taken from 
reverberation mapping closely agrees with the distance at 
which the H$\beta$ line peaks in the simulated line emissivity profile.
\item The NLR from our disk-like model is denser as it is postulated from observations. NLR clouds 
may become rare while escaping from accretion disk atmosphere, which we plan to take into account in the future paper. 

\end{enumerate}

\acknowledgments
{\small This research was supported by Polish National Science 
Center grants No. 2015/17/B/ST9/03422, 2015/18/M/ST9/00541, 2015/17/B/ST9/03436,
2016/21/N/ST9/03311 and 
by Ministry of Science and Higher Education grant W30/7.PR/2013.
It received funding from the European Union Seventh Framework Program 
(FP7/2007-2013) under the grant agreement No.312789.
TPA received funding from NCAC PAS grant for
young researchers. GJF thanks the Nicolaus Copernicus Astronomical Center for its hospitality 
and acknowledges support by NSF (1108928, 1109061, and 1412155), 
NASA (10-ATP10-0053, 10-ADAP10-0073, NNX12AH73G, and ATP13-0153), 
and STScI (HST-AR- 13245, GO-12560, HST-GO-12309, GO-13310.002-A, 
HST-AR-13914, and HST-AR-14286.001).
}
\software{{\sc cloudy} (v17.00; \citep{Ferland2017})}



\bibliographystyle{apj}
\bibliography{refs}
\end{document}